\documentclass[pra,aps,twocolumn,floatfix]{revtex4}
\usepackage{graphicx}
\usepackage{amsmath}    
\usepackage{amsfonts}   
\usepackage{amssymb}    
\usepackage{amsbsy} 	
\usepackage{hyperref}
\usepackage{epstopdf}
\usepackage{lipsum}
\begin{document}
	
	\title{Carrier-Envelope-Phase Dependent Strong-Field Excitation}
	
	\author{D.~Chetty$^1$} 
	\author{R.~D.~Glover$^{1,2}$}
	\author{X.~M.~Tong$^{3}$}
	\author{B.~A.~deHarak$^{1,4}$}
	\author{H.~Xu$^1$}
	\author{N.~Haram$^1$}
	\author{K.~Bartschat$^{5}$}
	\author{A.~J.~Palmer$^1$}
	\author{A.~N.~Luiten$^2$}
	\author{P.~S.~Light$^2$}
	\author{I.~V.~Litvinyuk$^1$}
	\author{R.~T.~Sang$^1$}	
	\email{r.sang@griffith.edu.au}
	\affiliation{$^1$Centre for Quantum Dynamics, Griffith University, Brisbane, QLD 4111, Australia}
	\affiliation{$^2$Institute for Photonics and Advanced Sensing and School of Physical Sciences, The University of Adelaide, Adelaide, SA 5005, Australia}
	\affiliation{$^3$Center for Computational Sciences, University of Tsukuba, 1-1-1 Tennodai, Tsukuba, Ibaraki 305-8573, Japan}
	\affiliation{$^4$Physics Department, Illinois Wesleyan University, Bloomington, IL 61702-2900, USA}
	\affiliation{$^5$Department of Physics and Astronomy, Drake University, Des Moines, IA 50311, USA}

\date{\today}
	
\begin{abstract}
We present a joint experimental-theoretical study on the effect of the carrier-envelope phase (CEP) of a few-cycle pulse on the atomic excitation process. 
We focus on the excitation rates of argon as a function of CEP in the intensity range from 50$-$300 TW/cm$^2$, which covers the transition between the multi\-photon and tunneling regimes.
Through numerical simulations based on solving the time-dependent Schr\"{o}dinger equation (TDSE), we show that the resulting bound-state population is highly sensitive to both the intensity and the CEP. 
Because the intensity varies over the interaction region, the CEP effect is considerably reduced in the experiment. 
Nevertheless, the data clearly agree with the theoretical prediction, and the results encourage the use of precisely tailored laser fields to coherently control the strong-field excitation process.
We find a markedly different behavior for the CEP-dependent bound-state population at low and high intensities with a clear boundary, which we attribute to the transition from the multi\-photon to the tunneling regime.
 
 
\end{abstract}

\maketitle

High-intensity lasers provide access to highly excited bound states that have a variety of applications. 
For example, excited states that decay directly to the ground state produce coherent EUV light through below-threshold harmonic (BTH) generation~\cite{Xiong2016,Zhao2019}, 
and in noble gases long-lived metastable states may be populated, either directly or through cascade decay of higher states~\cite{NubPRL08,Chetty2020}. 
This offers an alternative excitation pathway for a variety of applications~\cite{Baker2004,Lu2014,Sturchio2014,Knecht2013,Ohayon2018}.
In this commonly referred to ``strong-field regime'', excitation mechanisms are typically explained using either the multi\-photon (MP) or tunneling picture, 
with the Keldysh parameter $\gamma$~\cite{Keldysh64} providing a measure for which one is most appropriate.
In the MP regime ($\gamma \gg 1$), an excited state can be reached via the absorption of multiple photons whose energy add up. 
In the tunneling regime ($\gamma < 1$), excitation is the result of recapturing tunneled electrons, a process usually referred to as frustrated tunnel ionization (FTI)~\cite{NubPRL08}.
In the intermediate regime ($\gamma \approx 1$), there is a rich variety of physics, as there are contributions from both MP and tunneling effects. 

For few-cycle pulses, the carrier-envelope phase (CEP) becomes an important parameter for controlling interactions, for example, the CEP has been shown to play a 
crucial role in processes such as high-harmonic generation~\cite{Christov97}, above-threshold ionization~\cite{Haworth2007}, 
generation of attosecond pulses~\cite{Kienberger2002,Baltuvska2003}, coherent control of molecular dynamics~\cite{Bhardwaj2001,Han2014}, and control of BTH~\cite{Chin14,Yun18}. 
Because excitation is described very differently in the MP and tunneling regimes, it is expected that the effect of changing the CEP will manifest 
differently depending on the intensity.  There are only a few studies to date that have explored how the CEP affects the final bound-state populations, particularly in the tunneling regime. 
The CEP can potentially be used to control populations or serve as a clear marker for the changing dynamics of the interaction from MP absorption to tunneling plus re-capture.

In this Letter, we analyze the effect of the CEP on strong-field excitation. 
We experimentally investigate excitation rates of argon as a function of CEP in the tunneling regime, where we find good agreement between our experimental data and numerical results based on the time-dependent Schr\"{o}dinger equation (TDSE). 
We show that the bound-state population strongly depends on the CEP, especially in the tunneling regime where both the distribution of populated states and the total excitation rates are highly sensitive to the CEP. 
The TDSE results show that at lower laser intensities, intermediate between the MP and tunneling regimes, a remarkable change occurs in the dependence of the final bound-state populations on the CEP. 
The change in behavior indicates the transition of the dynamics from MP excitation to recaptured tunneling ionization. 

Excitation of atoms or molecules in the strong-field regime is unique. 
The laser field is comparable to the field strength between the outer electron(s) and the nucleus, and there are many different excitation pathways to a complex manifold of excited states. 
Furthermore, the energy levels are strongly modified due to the AC Stark shift, and the excitation probability to a particular target state will depend strongly on the intensity of the laser field. 
The laser intensity can be used to fine-tune excitation in a similar way the frequency is used to tune interactions in weak fields, known as the resonant dynamic Stark shift~\cite{Bucksbaum2004,Rabitz2000,Bunjac2017}. 
This provides access to states that are usually inaccessible and can be used to enhance ionization yields via resonantly enhanced MP ionization~\cite{Wollenhaupt2013}.

Regularly spaced enhancements exist that depend on the peak laser intensity~\cite{Chetty2020,ZimPRL17,Li2014JPhyB,Li2014,Piraux2017,Xu2020}. 
In the MP picture, these enhancements occur when the AC Stark-shifted ionization 
threshold crosses the energy level of an integer number $N$ of absorbed photons, known as channel closings~\cite{Kruit1983,Muller1983}. 
Consequently, high-energy Rydberg states come into resonance at regularly spaced intensity intervals 
with $\Delta U_P(I) = \omega$, where $U_P = I/4\omega^2$ is the ponderomotive energy of the electron, 
$I$ is the laser intensity, and $\omega$ is the laser frequency. [Unless indicated otherwise, we use atomic units throughout.]
At higher intensities, where tunnel ionization dominates, excitation is more commonly described in the time domain. 
In this picture, an electron tunnels through the distorted barrier, and the combined laser and Coulomb fields drive the 
electron dynamics. Under certain conditions, the laser field imparts negligible drift energy to the electron, and the 
Coulomb field recaptures it into an excited state at the conclusion of the pulse.  
Here, enhancements in excitation rates are a result of constructive inter\-ference between electron wavepackets emitted 
at subsequent field-cycle maxima, which are recaptured into the same quantum state~\cite{Hu2019,Xu2020}. 
In the strong-field approximation for sufficiently long pulses, electrons born one period~$T$ apart are launched in 
the same direction, and their subsequent dynamics are very similar. Their contributions to a particular state, however, differ by a 
phase \hbox{$\Delta\theta = T(U_P + I_P + E_n)$}. This leads to the condition for constructive interference as
\begin{equation}\label{eqInterference}
    m\omega = I_P + E_{nl} + U_P,
\end{equation}
where $m$ is a real integer, $E_{nl}$ is the field-free energy of the $nl$ orbital, and $I_P$ is the static ionization potential. 
This closely resembles the condition in the MP regime with the same intensity interval between successive enhancements.  

Few-cycle pulses contain a broad range of photon energies. We have previously shown that enhancements are broadened for CEP-averaged few-cycle pulses~\cite{Chetty2020}. 
In the MP picture, many resonance pathways exist, and thus an individual state may be reached through absorption of photons from any part of the frequency spectrum. 
Each pathway can be made up of a unique~$N$ that typically obeys the dipole selection rules~\cite{Chen2006,Arbo2008,Krajewska2012,Li2014,Piraux2017}. 
If we consider the contribution due to non-dipole transitions~\cite{Ni2020}, it does not change the conclusions here since the non-dipole effect is negligibly small.
The overall transition amplitude to a particular state is the coherent sum of the contributions from each pathway.
In this picture, the CEP has been found to have an effect under certain conditions~\cite{Nakajima2006PRL,Nakajima06,Peng2010,Jha2014}. 
When an excited state is located off-resonance, the excitation pathway may go through absorption of photons of different energy. 
Because the CEP dictates the phase difference between the contributions from different wavelengths, there is interference, and the CEP was found to modify the population. 
However, since this effect is only significant for off-resonant excitation, the CEP tends to modify only the population of states that have a very low excitation probability~\cite{Nakajima2006PRL,Nakajima06}.

In the tunneling picture, the CEP significantly alters the time dependence of the electric field.  Consequently, the dynamics of 
electrons born at subsequent field-cycle maxima are no longer similar. In this case, the electron wavepackets 
are born at different $U_P$, dependent on the CEP, and the condition for constructive interference given by 
Eq.~(\ref{eqInterference}) needs to be modified accordingly. 
Hence the energy of the final state where constructive interference occurs will be modified, and it is expected 
that modifying the CEP in turn modifies the relative populations of the resulting bound states. 
This has been demonstrated indirectly through BTH~\cite{Chin14,Yun18,Xiong2016}, but has yet to be shown 
directly either experimentally or theoretically.

\begin{figure}
    \centering
    \includegraphics[width=\linewidth]{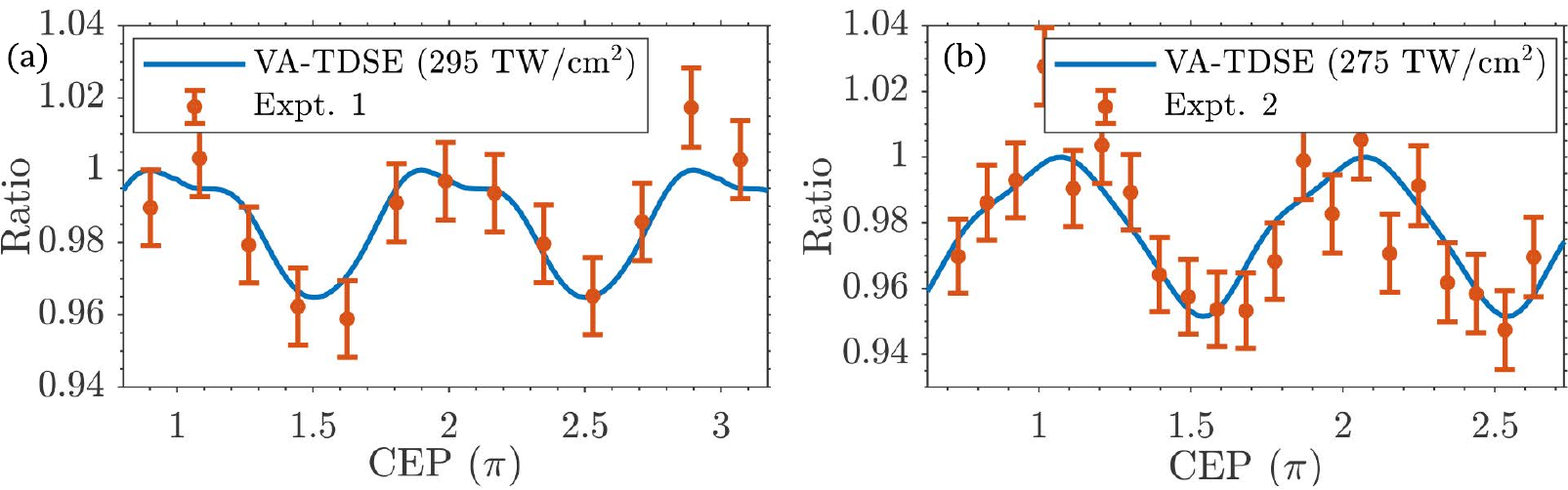}
    \caption{Comparison of the experimentally measured ratio of excitation yields to ionization yields and the volume-averaged
     TDSE results for two data sets. The absolute phase (unknown, see text) and normalized yield obtained in the experiment 
     were shifted to best fit theory. }
    \label{ThVsExp}
\end{figure}

In our experiment, we measure excitation yields of argon resulting from the interaction with CEP-stabilized few-cycle 
laser pulses at peak intensities of 275 and 295~TW/cm$^2$. The experimental procedure and detection methods are described in Ref.~\cite{Chetty2020}. 
Briefly, CEP-stabilized laser pulses are obtained from a commercial laser system (Femto Power Pro CE-Phase) with a pulse 
duration of $\sim$6~fs (FWHM) centered at 800~nm. 
A commercial $f-2f$ interferometer (Menlo APS800) located close to the interaction region is integrated to compensate 
long-term drifts and measure the stability of the locking system. 
The CEP is locked to a fixed (though arbitrary, i.e., not absolutely known) value, 
while the relative shift is controlled by translation of a fused-silica 
wedge (1~mm $\approx$ 1.25~rad phase shift). 
To quantify excitation rates after interaction with the laser, we measure the yields of excited atoms (Ar$^*$) surviving a $0.15-0.6$~ms flight time to a micro-channel plate detector. 
Simultaneously, we detect ionization yields (Ar$^+$) using a time-of-flight apparatus to distinguish charged and neutral particles.

Two results of these measurements together with numerical simulations are shown in Fig.~\ref{ThVsExp}. Because the measurement requires that the experimental 
parameters remain stable, we are restricted to high intensities where the integration times are small enough to maintain a 
stable CEP lock. At peak intensities below $\sim$250~TW/cm$^2$, long-term instability of the laser intensity dominates,
and we are unable to resolve a CEP effect. We observe the ratio of Ar$^*$/Ar$^+$, as this removes experimental systematics 
such as the density of the atomic beam, which has a small but significant fluctuation.
In both experiments, a clear modulation in the yields is observed with a similar level of modulation in the order 
of 5\% and a period of $\pi$, as required by symmetry of the excitation process. 

The comparison of the experimental measurements and the TDSE predictions (solid lines) shows very good 
agreement and gives us confidence that the numerical results are a good representation of the interaction process. 
The experimentally measured yields are affected by intensity-volume averaging (VA) and CEP fluctuations, and the numerical results 
have been processed to account for these. 
The details of the calculations have been described elsewhere~\cite{Chetty2020}. Briefly, they are based on solving 
the TDSE in the single-active-electron approximation by splitting the wavefunction into an inner and outer region. 
The outer wavefunction is projected onto momentum space, and the total ionization probabilities are obtained by 
integrating the electron momentum distribution over the entire momentum space. Quantum-state populations up to $n=30,l=29$ 
are found by projecting the inner-region wave function on the field-free atomic excited states. 
The total excitation probability, $P({\rm Ar}^*)$ is obtained by summing up all these populations.

\begin{figure}
    \centering
    \includegraphics[width=\linewidth]{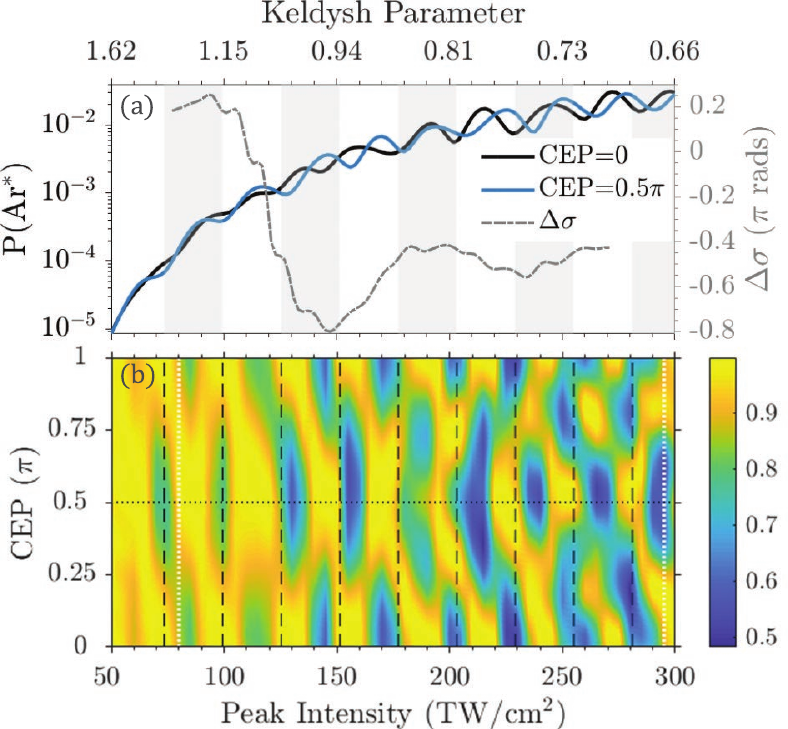}
    \caption{(a) Total excitation probabilities from TDSE calculations as a function of intensity for a cosine and sine few-cycle pulse. 
    The dashed line (referring to the right $y$-scale) shows the relative phase difference of the observed oscillations. 
    (b)~Excitation probabilities as a function of CEP 
    and intensity normalized to their respective maximum at each intensity. 
    The dotted  white lines correspond to lineouts shown in Fig.~\ref{PES}. 
    The black dashed lines mark the 13- to 22-photon channel closings and the shaded regions correspond to odd-photon absorption channels. No volume-averaging was performed.
    }
    \label{TDSE-norm-1}
\end{figure}

\begin{figure}
    \centering
    \includegraphics[width=\linewidth]{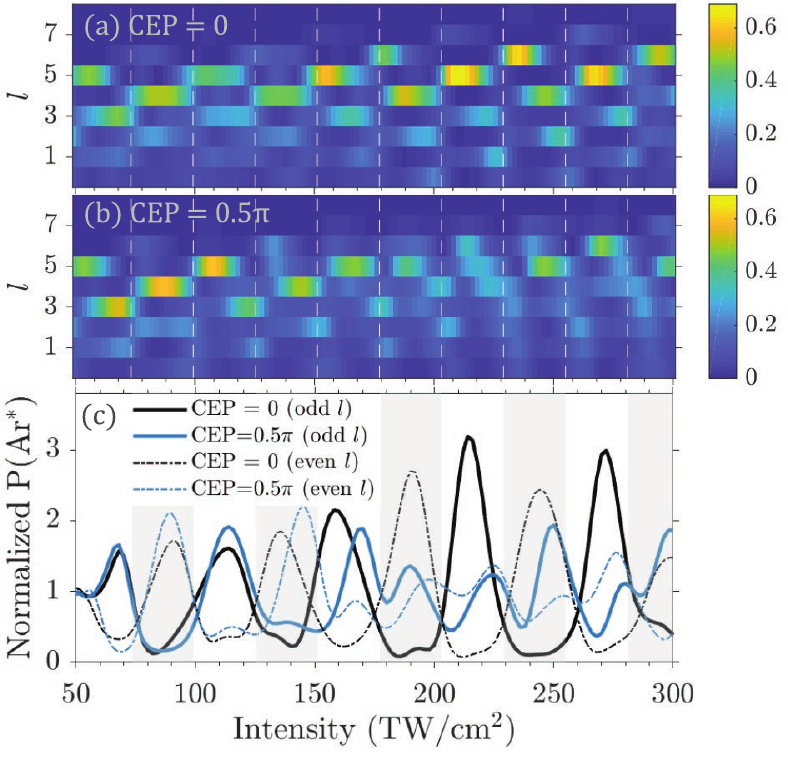}
    \caption{
    (a) Normalized probability for excitation to different $l$ states for a $\rm CEP=0$ (cosine) pulse.
    (b)~As in (a) but for a $\rm CEP=0.5\pi$ (sine) pulse.
    (c)~ Normalized probability for excitation to the sum of states with odd $l$ (thick solid lines) and even $l$ (thin dashed lines). 
    Thewhite dashed lines mark the 13- to 22-photon channel closings and the shaded regions correspond to odd-photon absorption channels. No volume-averaging was performed.
    }
    \label{TDSE-norm-2}
\end{figure}

We use the TDSE results to extend our study to lower intensities and compare the bound-state populations in the transition between the MP and tunneling regimes where we observe a clear change in behavior.
Our previous work~\cite{Chetty2020}, which measured the bound-state yields over a broad intensity range, demonstrates that our theory is valid at the intensities investigated here.
Figure~\ref{TDSE-norm-1}a compares the total excitation probability for a cosine and sine pulse ($\rm CEP=0$ and 0.5$\pi$, respectively). 
At low intensity, the excitation probability for the sine pulse maximizes at a slightly lower intensity (in each channel-closing cycle) than for the cosine pulse. 
A clear change in this behavior occurs at $\sim~125$~TW/cm$^2$ ($\gamma\sim 1 $), where the order in which the peaks are observed is flipped and the intensity separation between them is increased to the point 
where the two curves become almost completely out of phase. 
This is clearly shown in the plot by the relative phase difference of the oscillations, 
$\Delta\sigma$, which exhibits a sharp change starting at $\sim$100~TW/cm$^2$ and eventually experiences a shift of close to $2\pi/3$. 
We define the relative phase of the oscillations by first isolating the oscillations in $P$(Ar$^*$) for both the sine and cosine pulses by dividing $P$(Ar$^*$) by the smoothed CEP-averaged data. 
A moving-window FFT is used to extract the phase information for the highest amplitude oscillation frequency, and the difference between phases $\Delta\sigma$ for the sine and cosine pulses is plotted.
The clear modification in behavior in this region indicates the changing dynamics from the multiphoton to the tunneling regime.
In the MP regime, the sine pulse leading the cosine pulse is consistent with observations of CEP effects using few-cycle rf-pulses~\cite{Li2010CEP} and is due to the interference between different MP pathways. 
In the tunneling regime, the ponderomotive potential at the maximum of the field cycle is larger for the cosine pulse. 
Hence, the condition for constructive interference given in Eq.~(\ref{eqInterference}) is reached at a lower intensity compared to the sine pulse, and the cosine pulse is observed to lead the sine pulse.

At the same intensity where the relative phase change is observed, we also notice an increase in the modulation depth when changing the CEP. 
Figure~\ref{TDSE-norm-1}(b) shows the excitation probabilities for each intensity scaled to the maximum excitation probability at that intensity for all CEPs. 
Similar patterns repeat at a period close to that of the channel closings and this pattern continues from the low-intensity to the 
high-intensity regime, where the conditions for both channel closings and the quantum trajectory model used to derive Eq.(~\ref{eqInterference}) explain this periodicity. 
It  is  clear  that  the  sensitivity of the bound-state populations to the CEP is dependent on the intensity. Below the 15-photon 
channel closing ($\sim$125 TW/cm$^2$, $\gamma\sim 1$), the change in total excitation probability as a function of the CEP is approximately 20\%. 
At higher intensities, there is a step-like increase to $>40$\%, consistently approaching 50\% in successive channels.

The distribution of $l$ states after excitation has been studied previously both semi-classically~\cite{Arbo2008} and quantum mechanically~\cite{Hu2019,Li2014,Piraux2017,Chen2006}. 
In the MP picture excitation follows the dipole selection rules and for an argon atom, absorption of an even (odd) number of photons will preferentially 
populate odd (even) $l$'s, which is referred to as odd (even) parity. A change in parity between odd and even is indicative of an additional absorbed photon. 
Previously we showed that the parity effect is observed for CEP-averaged pulses~\cite{Chetty2020}.
Here we investigate whether this is true for both sine and cosine pulses. 
The distribution of $l$ states due to excitation from a cosine pulse is shown in figure~\ref{TDSE-norm-2}(a) and a sine pulse in figure~\ref{TDSE-norm-2}(b).
Figure~\ref{TDSE-norm-2}(c) shows the sum of the excitation probability over all odd and even $l$ states. 
At low intensity, the parity effect is observed for both CEPs with excitation to odd (even) $l$ occurring when an even (odd) number of photons are absorbed, consistent with the expectation from the MP model~\cite{Arbo2008,Venzke2018}.

For cosine pulses, we observe parity at all intensities investigated, with excitation to odd and even $l$'s at successive $n$-photon absorption channels. 
However, for sine pulses at intensities beyond $\gamma\sim1$, we show a clear loss of the parity effect. 
The population is spread across a range of quantum states with odd and even $l$'s populated in each $n$-photon absorption channel.
Venzke \textit{et al.}~\cite{Venzke2018} previously showed that the parity effect can break down with few-cycle pulses, where it was proposed to be a consequence of the bandwidth of the pulse. However, here we show that parity is observed in the low-intensity regime but depends on the CEP at higher intensities.
The fact that parity is observed for a few-cycle cosine pulse suggests that the bandwidth of the pulse is not directly responsible for the loss of the parity effect.
Thus, we attribute our observation to the break down of inversion symmetry of the electric field for a sine and cosine pulse. 
This is indicative of a transition in the excitation mechanism from the MP to the tunneling regime, since the spatial symmetry of the electric field should only matter 
in the tunneling-plus-recapturing excitation model. 

\begin{figure}
    \centering
    \includegraphics[width=\linewidth]{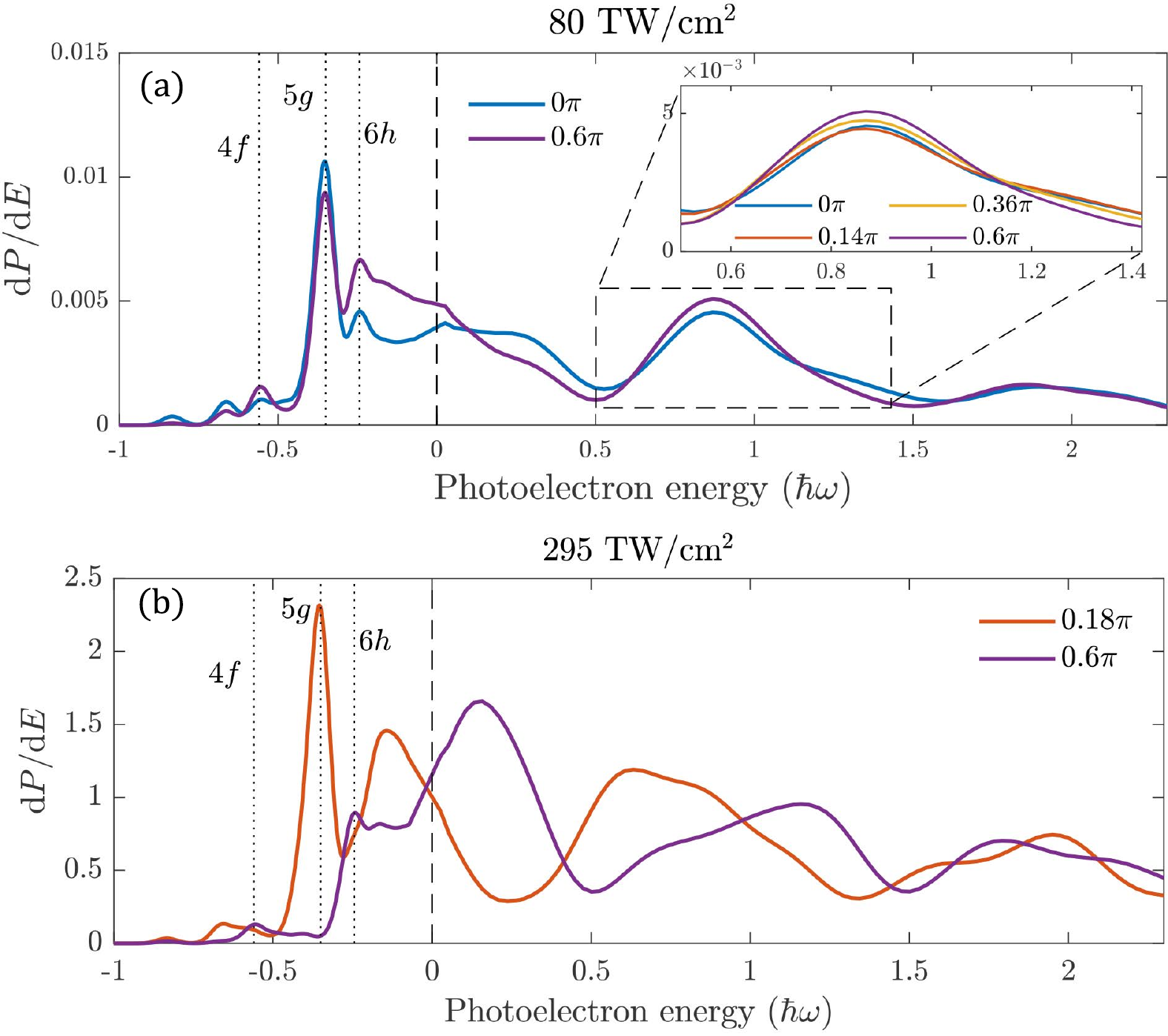}
    \caption{Normalized excitation and ionization spectra for select CEPs at peak intensities of (a) 80~TW/cm$^2$ and \hbox{(b)~295~TW/cm$^2$}. 
    The dashed line marks the shifted ionization potential, and the dotted lines mark the principal excited states.}
    \label{PES}
\end{figure}

Based on quantum defect theory~\cite{Seaton1983,Aymar1996,Gao2016}, the photo\-excitation 
and photo\-ionization processes can be treated in a unified way in the perturbative limit. Recently, 
Gao and Tong~\cite{Gao2019} showed that such a relation still exists even in a non\-perturbative regime. 
Figure~\ref{PES} shows the normalized photoelectron spectra across the ionization threshold 
(negative energy stands for excitation) at 80 and 295 TW/cm$^2$, which lie before and far after 
the step\-wise increase in modulation, respectively. We illustrate two CEPs that correspond to the minimum and maximum excitation yields. 
The lower intensity of 80~TW/cm$^2$ lies close to the 13-photon resonance to the $5g$ state. 
At this intensity, excitation to the $5g$ state and also to high-lying states near the continuum dominate the relative population. 
As the CEP is varied from 0 to 0.6$\pi$, there is a smooth change in the total number of bound-state electrons with a modulation depth of $\sim$15\%. 
When the total population is increased, the population of the $5g$ state is reduced. 
In fact, each state is observed to experience a unique CEP dependence. 
The modulation of the $5g$ state with CEP is significant, even near this resonance.
The change we observe in the structure of the bound-state spectrum is less significant in the low-energy photo\-electron spectrum, as shown in the inset of Fig.~\ref{PES}(a).

In contrast, at 295~TW/cm$^2$ there is a drastic change of the bound-state spectrum and also the continuum 
peak structure when the CEP is varied. 
At this high intensity, both the structure and the positions of the peaks are significantly 
modified with the CEP, accounting for the high sensitivity in bound-state populations observed in Fig.~\ref{TDSE-norm-1}(b). 
The populations of individual states are observed to be extremely CEP-sensitive, with the population 
of the $5g$ state fluctuating from a maximum to effectively zero with changing CEP.
However, comparing the CEP dependence at both intensities, one similarity is observed. 
It appears that maximum excitation occurs at a CEP that facilitates the transfer of near-zero momentum 
photoelectrons below the ionization threshold, resulting in a reduced number of small-momentum 
photoelectrons (cf.\ photoelectron populations near the ionization threshold in Fig.~\ref{PES}). 

The above similarities, together with the previous observations, indicate a transition in the excitation mechanism.
At intensities where tunneling rates are very small, MP excitation dominates and the CEP effect arises due to interference from pathways reached from the absorption of a different number of photons. 
This effect is pronounced for lowly-populated states reached through off-resonant paths but does not affect highly-populated states significantly~\cite{Nakajima2006PRL,Nakajima06,Peng2010}. 
Additionally, a parity effect is always observed regardless of the CEP. 
In the intensity regime where $\gamma$ approaches unity, excitation may occur through either direct MP absorption or recapture of tunneled electrons. 
These competing mechanisms complicate the overall CEP dependence. However a clear change in behavior is observed at 125~TW/cm$^2$.
MP excitation still occurs, but now the dynamics of the tunneled wavepacket begins to influence the bound-state population. 
As a result, selection rules are found to break down, and parity is not observed under all pulse parameters but is found to depend on the CEP.
Eventually, well into the tunneling regime, the phase dependence is dominated by the tunneling dynamics, where the CEP may 
influence the proportion of recaptured electrons through the modification of electron trajectories. 

\smallskip 
In conclusion, we experimentally demonstrated that bound-state populations are modified depending on the CEP of the driving laser pulse. 
This serves as a proof of principle that bound-state populations can be controlled through precisely engineered pulses. 
We hope that our findings will encourage the use of other means such as two-color fields, which can be used for precisely controlling the electric field~\cite{Han2021}. 
With such fields, it should be possible to coherently control bound-state populations in the same manner as via the CEP. 
We also predict that the CEP provides a unique method to explore the boundary between the MP and the tunneling regimes. 
Not only does the level of modulation of the excited-state populations increase in the tunneling regime, but we also observe more subtle effects: 
the change in relative phase of the peak structure in the intensity dependence and the loss of the parity effect for sine pulses. 
Both of these phenomena warrant further investigation with alternative experimental methods.

\smallskip 
This project is supported under the Australian Research Council's Linkage Infra\-structure, 
Equipment and Facilities scheme (project LE160100027).  
D.~Chetty is supported by an Australian Government RTP Scholarship. N.~Haram is supported by a Griffith University International Postgraduate
Research Scholarship (GUIPRS), and H.~Xu is supported by an ARC Discovery Early Career Researcher Grant No.\ DE130101628.
\hbox{X-.M.~T.}\ was supported by Multidisciplinary Cooperative Research Program in CCS, University of Tsukuba.
Further funding was provided by the United States National Science Foundation under grants 
No.\ PHY-1402899 and PHY-1708108 (BdH) as well as No.\ PHY-1803844 and No.\ PHY-2110023 (KB).
\bibliography{CEPBib}
\end{document}